\documentclass[11pt]{cernrep}
\usepackage{graphicx,wrapfig}

\def\Journal#1#2#3#4{{#1} {\bf #2} (#3) #4}

\def\NPB{{\em Nucl. Phys.}   {\bf B}}

\def\EJC{{\em Eur. Phys. J.} {\bf C}}
\def\CPC{\em Comp. Phys. Commun.}

\newcommand{\etal}{{\em et al.}}
\newcommand{\etjet}{\ensuremath{{E_T^{\rm jet}}}}
\newcommand{\etjetstar}{\ensuremath{{E_T^{\rm *,jet}}}}
\newcommand{\etajet}{\ensuremath{{\eta_{\rm jet}}}}
\newcommand{\etajetstar}{\ensuremath{{\eta^*_{\rm jet}}}}
\newcommand{\figref}[1]{\figurename~\ref{#1}}
\newcommand{\logxpomeron}{\ensuremath{{\log_{10}(x_\pom)}}}
\newcommand{\pom}{{I\!\!P}}
\newcommand{\ptjetone}{\ensuremath{{p_T^{\rm jet1}}}}
\newcommand{\xgamma}{\ensuremath{{x_\gamma}}}
\newcommand{\xgammajets}{\ensuremath{{x_\gamma^{\rm jets}}}}
\newcommand{\xpom}{\ensuremath{{x_\pom}}}
\newcommand{\zpomeronjets}{\ensuremath{{z_\pom^{\rm jets}}}}

\def\d{{\rm d}}
\def\p{I\!\!P}

\def\lr{\left( }
\def\rr{\right) }
\def\le{\left[ }
\def\re{\right] }

\def\beq{\begin{equation}}
\def\eeq{\end{equation}}
\def\bea{\begin{eqnarray}}
\def\eea{\end{eqnarray}}

\begin{document}
\title{Diffractive Dijet Production at HERA}
\author{A.\ Bruni$^1$, M.\ Klasen$^{2,3}$, G.\ Kramer$^3$ and S.\ Sch\"atzel$^4$}
\institute{
 $^1$ INFN Bologna, Via Irnerio 46, 40156 Bologna, Italy \\
 $^2$ Laboratoire de Physique Subatomique et de Cosmologie,
 Universit\'e Joseph Fourier/CNRS-IN2P3, 53 Avenue des Martyrs, 38026
 Grenoble, France \\
 $^3$ II.\ Inst.\ f\"ur Theoret.\ Physik, Universit\"at
 Hamburg, Luruper Chaussee 149, 22761 Hamburg, Germany \\
 $^4$ DESY FLC, Notkestr. 85, 22607 Hamburg, Germany}
\maketitle
\begin{abstract}
We present recent experimental data from the H1 and ZEUS Collaborations at
HERA for diffractive dijet production in deep-inelastic scattering (DIS) and
photoproduction and compare them with next-to-leading order (NLO) QCD
predictions using diffractive parton densities. While good agreement is
found for DIS, the dijet photoproduction data are overestimated by the NLO
theory, showing that factorization breaking occurs at this order. While this
is expected theoretically for resolved photoproduction, the fact that the
data are better described by a global suppression of direct {\em and}
resolved contribution by about a factor of two comes as a surprise. We
therefore discuss in some detail the factorization scheme and scale
dependence between direct and resolved contributions and propose a new
factorization scheme for diffractive dijet photoproduction.
\end{abstract}

\section{Introduction}
\label{sec:1}

\begin{figure}
 \centering
 \includegraphics[width=0.5\columnwidth]{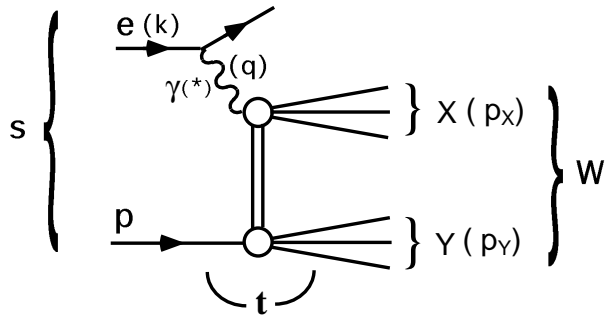}
 \caption{\label{fig:1}Diffractive scattering process $ep\to eXY$, where
 the hadronic systems $X$ and $Y$ are separated by the largest rapidity
 gap in the final state.}
\end{figure}

It is well known that in high-energy deep-inelastic $ep$-collisions a large
fraction of the observed events are diffractive. These events are defined
experimentally by the presence of a forward-going system $Y$ with
four-momentum $p_Y$, low mass $M_Y$ (in most cases a single proton and/or
low-lying nucleon resonances), small momentum transfer squared
$t=(p-p_Y)^2$, and small longitudinal momentum transfer fraction
$x_{\p}=q(p-p_Y)/qp$ from the incoming proton with four-momentum $p$ to the
system $X$ (see Fig.\ \ref{fig:1}). The presence of a hard scale, as for
%
%
example the photon virtuality $Q^2=-q^2$ in deep-inelastic scattering (DIS)
or the large transverse jet momentum $p_T^{*}$ in the photon-proton
centre-of-momentum frame, should then allow for calculations of the
production cross section for the central system $X$ with the known methods
of perturbative QCD. Under this assumption, the cross section for the
inclusive production of two jets, $e+p \rightarrow e+2~{\rm jets}+X'+Y$,
can be predicted from the  well-known formul\ae\ for jet production in
non-diffractive $ep$ collisions, where in the convolution of the partonic
cross section with the parton distribution functions (PDFs) of the proton
the latter ones are replaced by the diffractive PDFs. In the simplest
approximation, they are described by the exchange of a single, factorizable
pomeron/Regge-pole. \\

The diffractive PDFs have been determined by the H1 Collaboration at HERA
from high-precision inclusive measurements of the DIS process $ep
\rightarrow eXY$ using the usual DGLAP evolution equations in leading
order (LO) and next-to-leading order (NLO) and the well-known formula for
the inclusive cross section as a convolution of the inclusive parton-level
cross section with the diffractive PDFs \cite{h1ichep02}. For a similar
analysis of the inclusive measurements of the ZEUS Collaboration 
see~\cite{Chekanov:2004hy,zeus_mx}.  
A longer discussion of the extraction of diffractive
PDFs can also be found in these proceedings~\cite{Schilling}
and in~\cite{Martin:2004xw}. 
For inclusive diffractive DIS it has been proven
by Collins that the formula referred to above is applicable without
additional corrections and that the inclusive jet production cross section
for large $Q^2$ can be calculated in terms of the same diffractive PDFs
\cite{Collins:1997sr}. The proof of this factorization formula, usually
referred to as the validity of QCD factorization in hard diffraction, 
may be expected to hold
for the direct part of photoproduction ($Q^2\simeq0$) or
low-$Q^2$ electroproduction of jets \cite{Collins:1997sr}. However,
factorization does not hold for hard processes in diffractive hadron-hadron
scattering. The problem is that soft interactions between the ingoing two
hadrons and their remnants occur in both the initial and final state. This
agrees with experimental measurements at the Tevatron \cite{Affolder:2000vb}.
Predictions of diffractive dijet cross sections for $p\bar{p}$ collisions as
measured by CDF using the same PDFs as determined by H1 \cite{h1ichep02}
overestimate the measured cross section by up to an order of magnitude
\cite{Affolder:2000vb}. This suppression of the CDF cross section can be
explained by considering the rescattering of the two incoming hadron beams
which, by creating additional hadrons, destroy the rapidity gap
\cite{Kaidalov:2001iz}.\\

Processes with real photons ($Q^2 \simeq 0$) or virtual photons with fixed,
but low $Q^2$ involve direct interactions of the photon with quarks from the
proton as well as resolved photon contributions, leading to parton-parton
interactions and an additional remnant jet coming from the photon (for a
review see \cite{Klasen:2002xb}). As already said, factorization should be
valid for direct interactions as in the case of DIS, whereas it is expected
to fail for the resolved process similar as in the hadron-hadron scattering
process. In a two-channel eikonal model similar to the one used to calculate
the suppression factor in hadron-hadron processes \cite{Kaidalov:2001iz},
introducing vector-meson dominated photon fluctuations, a suppression by
about a factor of three for resolved photoproduction at HERA is predicted
\cite{Kaidalov:2003xf}. Such a suppression factor has recently been applied
to diffractive dijet photoproduction \cite{Klasen:2004tz,Klasen:2004qr} and
compared to preliminary data from H1 \cite{h1ichep04} and ZEUS
\cite{zeusichep04}. While at LO no suppression of the resolved contribution
seemed to be necessary, the NLO corrections increase the cross section
significantly, showing that factorization breaking occurs at this order at
least for resolved photoproduction and that a suppression factor $R$ must
be applied to give a reasonable description of the experimental data. \\

As already mentioned elsewhere \cite{Klasen:2004tz,Klasen:2004qr},
describing the factorization breaking in hard photoproduction as well as in
electroproduction at very low $Q^2$ \cite{Klasen:2004ct} by suppressing the
resolved
contribution only may be problematic. An indication for this is the fact
that the separation between the direct and the resolved process is uniquely
defined only in LO. In NLO these two processes are related. The separation
depends on the factorization scheme and the factorization scale
$M_{\gamma}$. The sum of both cross sections is the only physically relevant
cross section, which is approximately independent of the factorization
scheme and scale \cite{BKS}. As demonstrated in Refs.\
\cite{Klasen:2004tz,Klasen:2004qr} multiplying the resolved cross
section with the suppression factor $R=0.34$ destroys the correlation of the
$M_{\gamma}$-dependence between the direct and resolved part,
and the sum of both parts has a stronger $M_{\gamma}$-dependence than for
the unsuppressed case ($R=1$), where the $M_{\gamma}$-dependence of the NLO
direct cross section is compensated to a high degree against the
$M_{\gamma}$-dependence of the LO resolved part. \\

In the second Section of this contribution, we present the current experimental
data from the H1 and ZEUS Collaborations on diffractive dijet production in
DIS and photoproduction and compare these data to theoretical predictions at
NLO for two different scenarios: suppression of only the resolved part by a
factor $R=0.34$ as expected from LO theory and proposed in
\cite{Kaidalov:2001iz}, and equal suppression of all direct and resolved
contributions by a factor $R=0.5$, which appears to describe the data better
phenomenologically. This motivates us to investigate in the third Section the
question whether certain parts of the direct contribution might break
factorization as well and therefore need a suppression factor. \\

The introduction of the
resolved cross section is dictated by perturbation theory. At NLO, collinear
singularities arise from the photon initial state, which are absorbed at the
factorization scale into the photon PDFs. This way the photon PDFs become
$M_{\gamma}$-dependent. The equivalent $M_{\gamma}$-dependence, just with
the opposite sign, is left in the NLO corrections to the direct
contribution. With this knowledge, it is obvious that we can obtain a
physical cross section at NLO, {\it i.e.} the superposition of the NLO
direct and LO resolved cross section, with a suppression factor $R<1$ and no
$M_{\gamma}$-dependence left, if we also multiply the
$\ln M_{\gamma}$-dependent term of the NLO correction to the direct
contribution with the same suppression factor as the resolved cross
section. We are thus led to the theoretical conclusion that, 
contrary to what one may expect,
not {\em all} parts of the direct contribution factorize. Instead, the
{\em initial state} singular part appearing beyond LO breaks factorization
even in direct photoproduction, presumably through soft gluon attachments
between the proton and the collinear quark-antiquark pair emerging from
the photon splitting. This would be in agreement with the general remarks
about initial state singularities in Ref.\ \cite{Collins:1997sr}. \\

In the third Section of this contribution, we present the special form of the
$\ln M_{\gamma}$-term in the NLO direct contribution and demonstrate that the
$M_{\gamma}$-dependence of the physical cross section cancels to a large
extent in the same way as in the unsuppressed case ($R=1$).
These studies can be done for photoproduction ($Q^2 \simeq 0$) as well
as for electroproduction with fixed, small $Q^2$. Since in electroproduction
the initial-state singularity in the limit $Q^2 \rightarrow 0$ is more
directly apparent than for the photoproduction case, we shall
consider in this contribution the low-$Q^2$ electroproduction case just for
demonstration. This diffractive dijet cross section has been calculated
recently \cite{Klasen:2004ct}.
A consistent factorization scheme for low-$Q^2$ virtual photoproduction
has been defined and the full (direct and resolved) NLO corrections for
inclusive dijet production have been calculated in \cite{Klasen:1997jm}. In
this work we adapt this inclusive NLO calculational framework to diffractive
dijet production at low-$Q^2$ in the same way as in \cite{Klasen:2004ct},
except that
we multiply the $\ln M_{\gamma}$-dependent terms as well as the resolved
contributions with the same suppression factor $R=0.34$, as an example, as in
our earlier work \cite{Klasen:2004tz,Klasen:2004qr,Klasen:2004ct}.
The exact value of this
suppression factor may change in the future, when better data for
photoproduction and low-$Q^2$ electroproduction have been analyzed.
We present the $\ln M_{\gamma}$-dependence of the partly
suppressed NLO direct and the fully suppressed NLO resolved cross section
$\d\sigma/\d Q^2$ and their sum for the lowest $Q^2$ bin, before we give
a short summary in section 4.

\section{Comparison of H1 and ZEUS Data with NLO Theory Predictions}
\label{sec:2}
\begin{wrapfigure}{r}{8cm}
\setlength{\unitlength}{1cm}
\vspace*{-1.3cm}
\begin{picture}(16,4)
\put(0,0){%
\includegraphics[width=4cm,keepaspectratio]{%
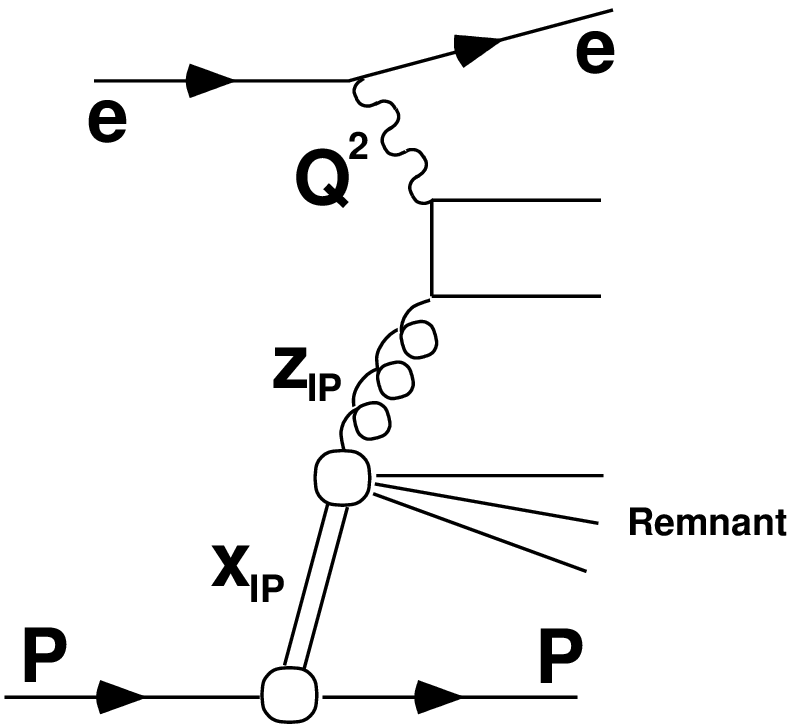}
}
\put(4,0){%
\includegraphics[width=4cm,keepaspectratio]{%
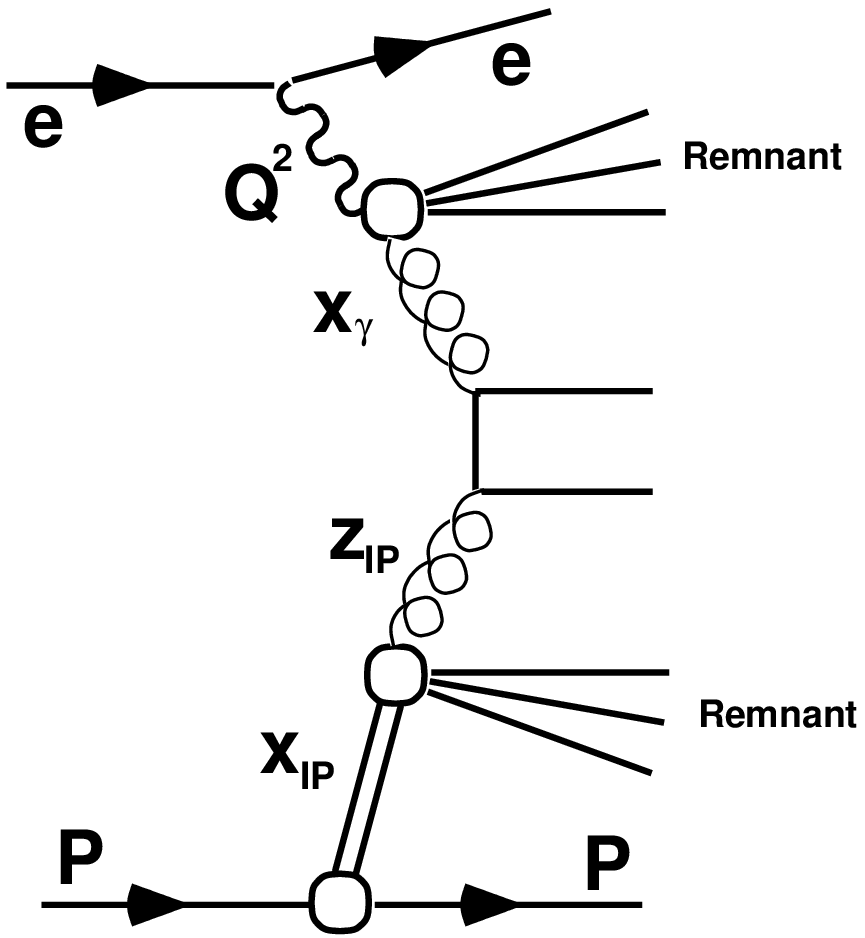}
}
\put(0,2) {\textbf{a)}}
\put(4,2) {\textbf{b)}}
\end{picture}
\vspace*{-0.8cm}
\caption{Example processes for a) direct photon and b)
 resolved photon interactions.}
\label{fig:bgf}
\vspace*{-0.3cm}
\end{wrapfigure}

In this Section, diffractive PDFs~\cite{h1ichep02,Chekanov:2004hy,zeus_mx} 
extracted from diffractive structure function data 
are used in NLO calculations to test factorisation in
diffractive dijet production. 
Dijet production is directly sensitive
to the diffractive gluon (\figref{fig:bgf}) whereas in inclusive
measurements the gluon is determined from scaling violations. 

\newpage

\subsection{Diffractive Dijet Production in DIS}
\begin{figure}[hhh]
\centering
\includegraphics[width=0.65\textwidth,keepaspectratio]{%
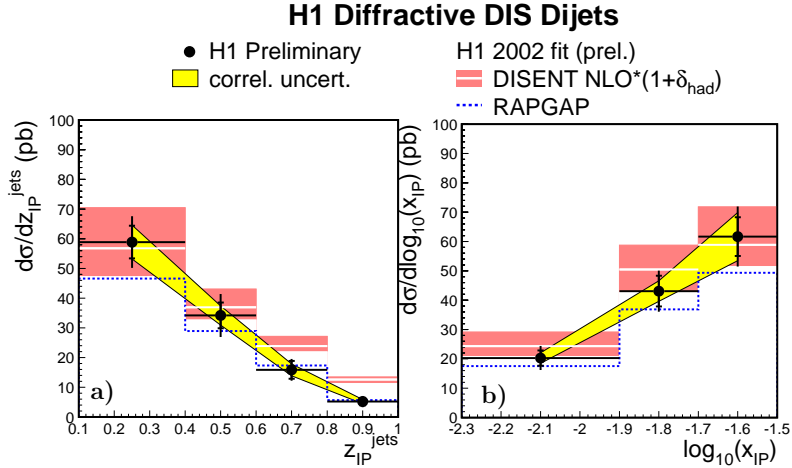}
\caption{Diffractive DIS dijet cross sections
compared with a NLO prediction based on diffractive PDFs and with RAPGAP.}
\label{dis2}
\end{figure}

H1 has measured the cross sections for dijet production~\cite{h1ichep04}
in the kinematic range $Q^2>4$~GeV$^2$, 
$165<W<242$ GeV (photon-proton centre-of-mass energy) and
$\xpom<0.03$.
Jets are identified using the inclusive $k_T$ cluster 
algorithm and selected by requiring $\etjetstar(1,2)>5,4$~GeV 
and $-3<\etajetstar<0$.\footnote{The '$*$' denotes variables in the photon-proton centre-of-mass system.}
NLO predictions have been obtained by interfacing the 
H1 diffractive PDFs with
the \mbox{DISENT} program~\cite{disent}.
The renormalisation and factorisation scales were set to the
transverse energy of the leading parton jet.
The NLO parton jet cross sections have been corrected for
hadronisation effects using the leading order (LO) generator RAPGAP~\cite{rapgap} with parton 
showers and the Lund string 
fragmentation model.
Comparisons of the DISENT and RAPGAP predictions with the measured
cross section differential in $\zpomeronjets$, an estimator for the
fraction
of the momentum of the diffractive exchange entering the hard scatter,
 are 
shown in \figref{dis2}a.
\begin{figure}[hhh]
\centering
\includegraphics[width=0.65\textwidth,keepaspectratio]{%
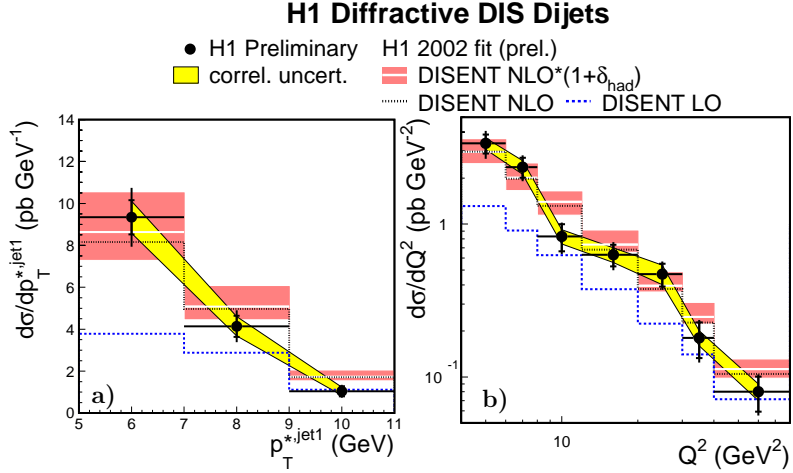}
\caption{Diffractive DIS dijet cross sections
compared with a NLO prediction based on diffractive PDFs.}
\label{dis}
\end{figure}
The inner band around the
NLO calculation indicates the $\approx 20\%$ uncertainty resulting
from a variation of
the renormalisation scale by factors 0.5 and 2. 
The uncertainty in the diffractive PDFs is not shown. Within this additional
uncertainty, which is large at high \zpomeronjets, the cross section
is well described.
The cross section differential in \logxpomeron{}, \ptjetone, and $Q^2$ is
shown in  \figref{dis2}b and \figref{dis}. All distributions are well
described and QCD factorisation is therefore in good
agreement with dijet production in diffractive DIS.

%
\begin{figure}[htb]
\centering
\includegraphics[width=0.65\textwidth,keepaspectratio]{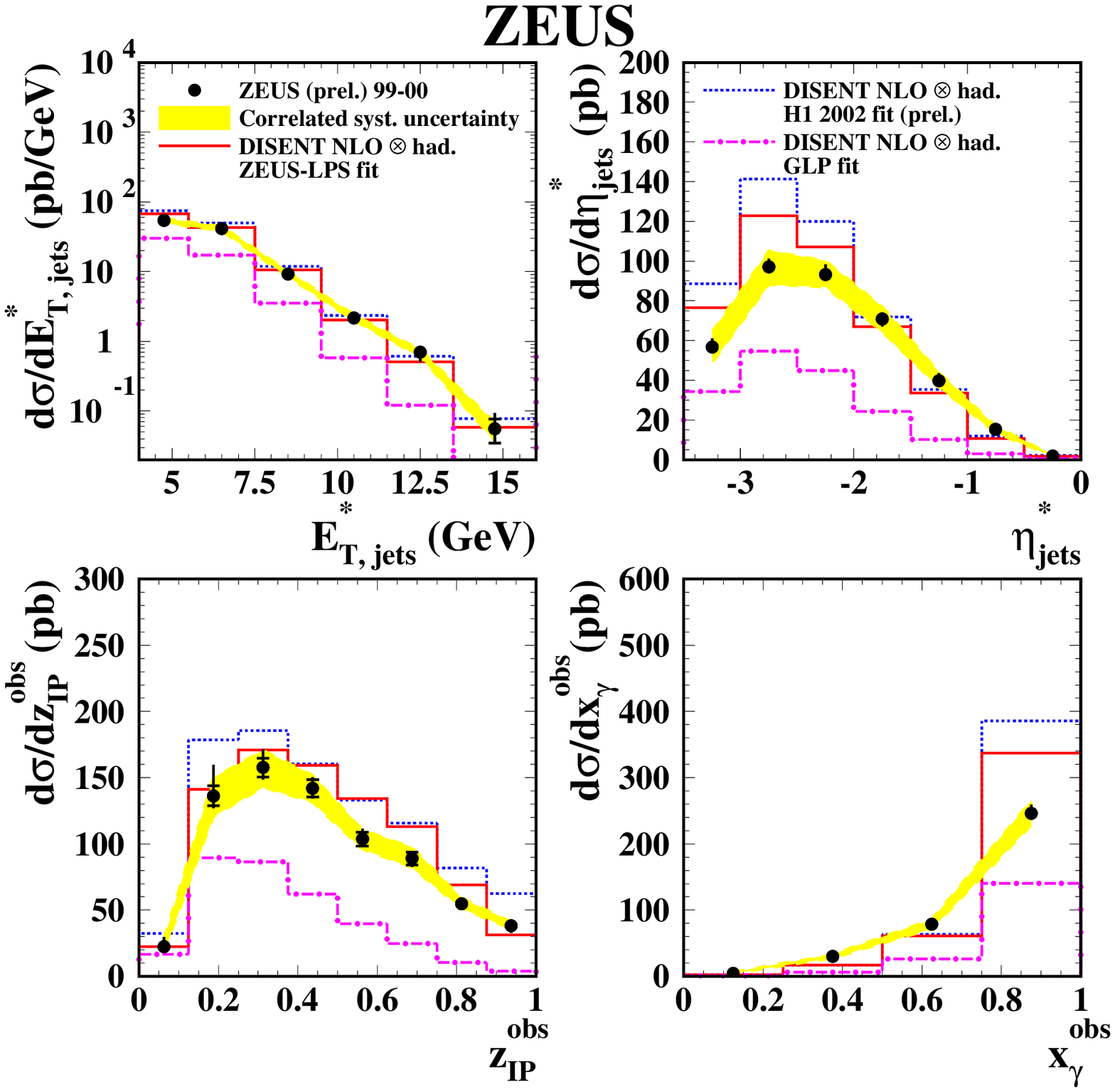}
\caption{Diffractive DIS dijet cross sections
compared with NLO predictions based on three sets of diffractive PDFs.}
\label{zeus_dis}
\end{figure}
Similar results are presented by ZEUS~\cite{zeus_disjets};
the dijet cross sections have been measured
in the kinematic range $5<Q^2<100$ GeV$^2$,
$100<W<200$ GeV, $\xpom<0.03$.
The jets were identified using the inclusive $k_T$ algorithm in the $\gamma p$
frame and required to satisfy $\etjetstar(1,2)>5,4$~GeV
and  $-3.5<\etajetstar<0.0$.
NLO predictions have been obtained with the \mbox{DISENT} program
interfaced to three different sets of diffractive PDFs:
from fit to H1 data~\cite{h1ichep02}, from fit to the ZEUS $M_X$ data (GLP)~\cite{zeus_mx}
and from fit to ZEUS LPS and $F_2^{D,charm}$ data~\cite{Chekanov:2004hy}.
Comparisons of the DISENT predictions with the measured
cross section differential in $\etjetstar$, $\etajetstar$, $\zpomeronjets$ and $x_{\gamma}^{obs}$
are shown in \figref{zeus_dis}. 
The $20-30\%$ uncertainty in the NLO calculations 
resulting from a variation of the renormalisation and factorisation scales is not shown.
Within the experimental and QCD scale uncertainties, the predictions
based on the H1 and ZEUS-LPS PDFs give a good description
of the dijet cross section.
The normalisation of the prediction using the GLP fit is substantially
lower than those from the other two sets of PDFs. 
For ZEUS, the difference observed between the three sets may be interpreted 
as an estimate of the uncertainty associated with the diffractive PDFs
and with the definition of the diffractive region.
The dijet data could be included in future fits in order to better constrain the diffractive
gluon distribution.

%
Within the experimental and theoretical uncertainties 
and assuming the H1 diffractive PDFs, factorisation is 
in good agreement with diffractive $D^*$ production~\cite{dstar,zeus_dstar} 
in the DIS kinematic region.

\subsection{Diffractive Photoproduction of Dijets}
\begin{figure}[hhh]
\centering
\includegraphics[width=0.65\textwidth,keepaspectratio]{%
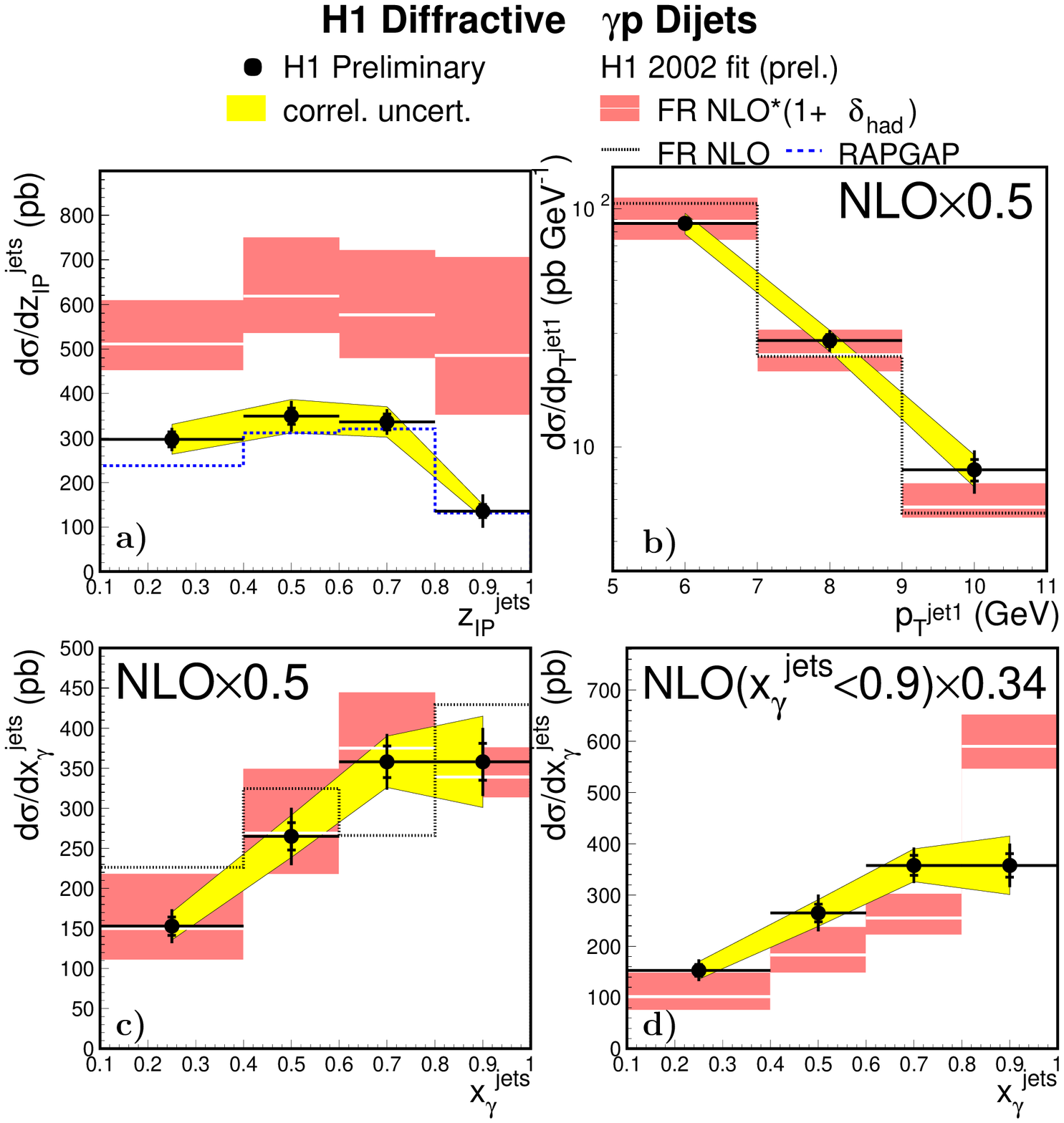}
\caption{a) Diffractive dijet photoproduction cross section differential
  in \zpomeronjets{} compared with a NLO prediction based on
  diffractive PDFs and RAPGAP. b)-d): Cross section differential in 
\ptjetone and \xgammajets, compared with the NLO prediction modified
  as follows: in b) and  c) the calculation is scaled by a global factor 0.5 whereas
in d) only the ``resolved'' part is scaled by 0.34.}
\label{gp}
\end{figure}

In photoproduction, a sizeable contribution to the cross section is
given by resolved photon processes
(\figref{fig:bgf}b) in which only a fraction $\xgamma <1$ of the
photon
momentum enters the hard scatter.
The photoproduction dijet cross section measured by H1 ($Q^2<0.01$~GeV$^2$,
$165<W<242$~GeV, $\xpom<0.03$, $\etjet(1,2)>5,4$~GeV, $-1<\etajet<2$,
inclusive $k_T$ algorithm)
is shown in \figref{gp}~\cite{h1ichep04}. 
NLO predictions have been obtained with the
Frixione \etal{} program\,\cite{frixione} interfaced to the
H1 diffractive PDFs. The parton jet calculation is corrected for hadronisation effects
using RAPGAP.
The cross section differential in \zpomeronjets{} is shown in
\figref{gp}a. The calculation lies a factor $\approx 2$ above the data.
\figref{gp}b and \ref{gp}c show the cross section as a function of $\ptjetone$
and $\xgammajets$ and the NLO predictions have been scaled down by a
factor 0.5. Good agreement is obtained for this global suppression.
In \figref{gp}d, only the ``resolved'' part for which $\xgammajets <
0.9$ at the parton level is scaled by a factor 0.34. This factor was proposed
by Kaidalov \etal\,\cite{Kaidalov:2003xf} for the suppression of the resolved
part in LO calculations. The calculation for
$\xgammajets>0.9$ is left unscaled. This approach is clearly
disfavoured.

%
\begin{figure}[hhh]
\centering
\includegraphics[width=0.65\textwidth,keepaspectratio]{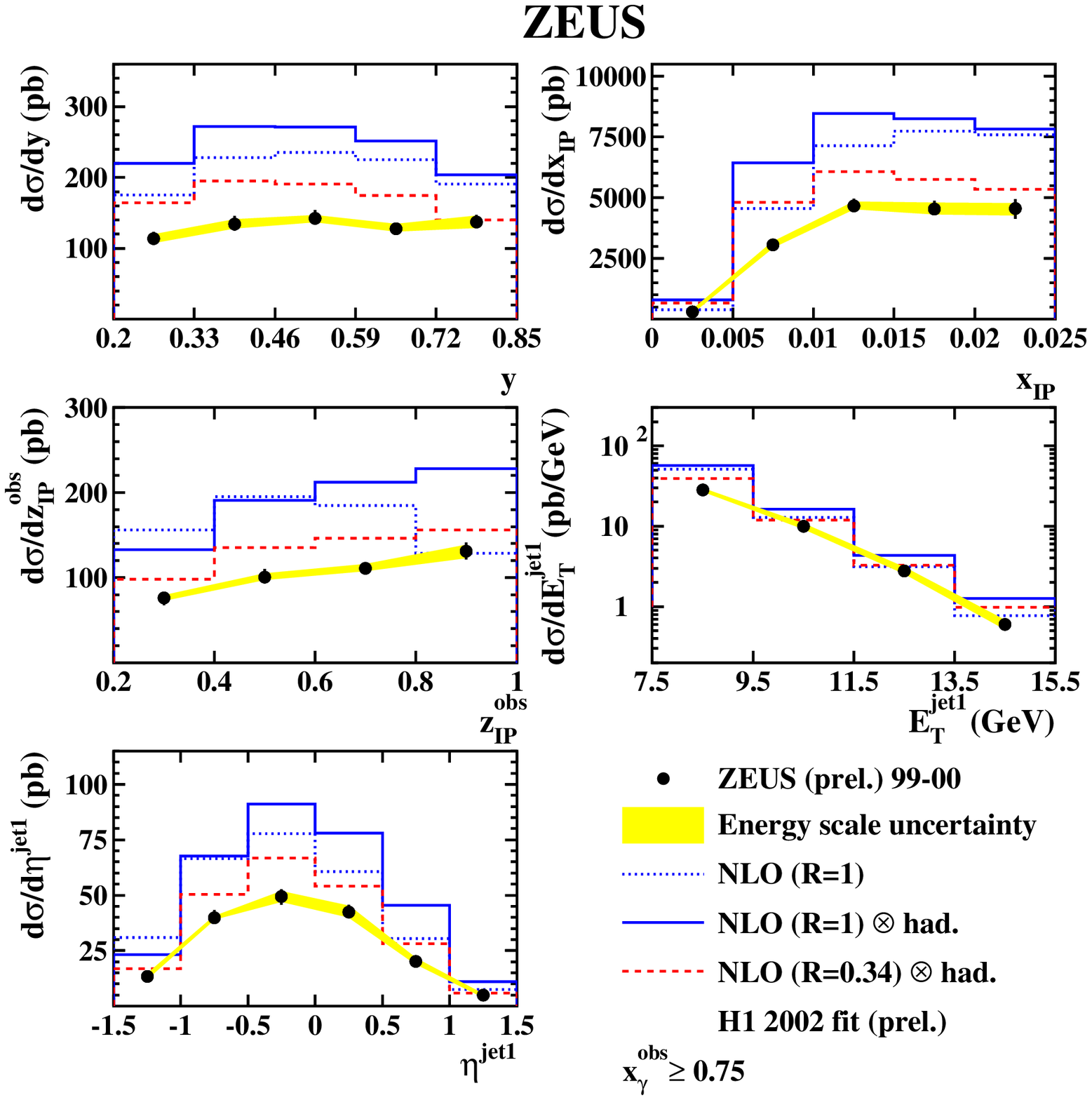}
\caption{Direct enriched photoproduction.
Diffractive dijet photoproduction cross section differential
in $y$, $\xpom$, $\zpomeronjets$, $\etjet^1$ and $\etajet_1$
compared with a NLO prediction based on diffractive PDFs. 
The NLO prediction is also presented 
corrected for hadronization effects and with the ``resolved'' part 
scaled by 0.34.}
\label{zeus_gp1}
\end{figure}
\begin{figure}[hhh]
\centering
\includegraphics[width=0.65\textwidth,keepaspectratio]{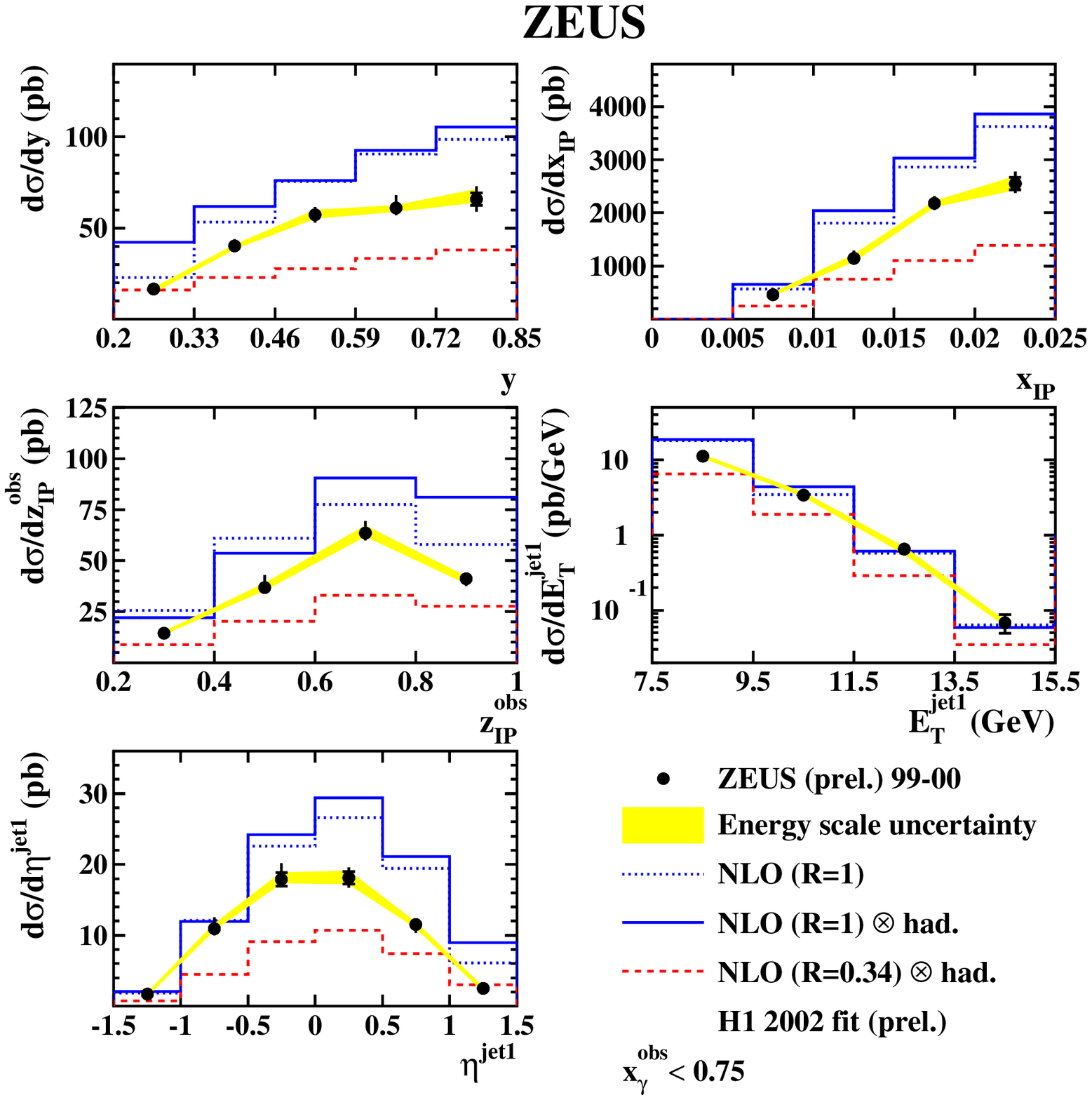}
\caption{ Resolved enriched photoproduction.
Diffractive dijet photoproduction cross section differential
in $y$, $\xpom$, $\zpomeronjets$, $\etjet^1$ and $\etajet_1$
compared with a NLO prediction based on diffractive PDFs. 
The NLO prediction is also presented 
corrected for hadronization effects and with the ``resolved'' part 
scaled by 0.34.}
\label{zeus_gp2}
\end{figure}
The ZEUS measurement~\cite{zeus_gp}
($Q^2<0.01$~GeV$^2$,
 $\xpom<0.025$, $0.2<y<0.85$, $\etjet(1,2)>7.5, 6.5$~GeV, $-1.5<\eta<1.5$, 
inclusive $k_T$ algorithm)
is shown in~\figref{zeus_gp1} and~\figref{zeus_gp2} 
separately for samples enriched in ``direct'' ($\xgammajets>0.75$) 
and ``resolved'' ($\xgammajets<0.75$) processes, 
respectively. 
The NLO~\cite{Klasen:2004qr} prediction using the H1 diffractive 
PDFs is also presented 
corrected for hadronization effects and with the ``resolved'' part 
scaled by the factor 0.34.
No evidence is observed for a suppression of resolved photon processes
relative to direct photon processes in any particular kinematic region.

Diffractive dijet photoproduction is overestimated by calculations
based on PDFs which give a good description of the diffractive DIS data.
Factorisation is broken in photoproduction relative to DIS by a factor 
$\approx 0.5$ with no observed dependence on $\xgamma$
or other kinematic variables.
%


\newpage

\section{Factorization and its Breaking in Diffractive Dijet Production}
\label{sec:3}

The fact that equal suppression of
direct {\em and} resolved photoproduction by a factor $R=0.5$ appears to
describe the H1 and ZEUS data better phenomenologically motivates us to
investigate in some detail the question whether certain parts of the direct
contribution might break factorization as well and therefore need a
suppression factor. These studies can be done for photoproduction ($Q^2
\simeq 0$) as well as for electroproduction with fixed, small $Q^2$. Since
in electroproduction the initial-state singularity in the limit $Q^2
\rightarrow 0$ is more directly apparent than for the photoproduction case,
we shall consider in this contribution the low-$Q^2$ electroproduction case
just for demonstration. \\

A factorization scheme for virtual photoproduction has been defined and the
full NLO corrections for inclusive dijet production have been calculated in
\cite{Klasen:1997jm}. They have been implemented in the NLO Monte Carlo
program JET\-VIP \cite{Potter:1999gg} and adapted to diffractive dijet
production in \cite{Klasen:2004ct}. The subtraction term, which is absorbed
into the PDFs of the virtual photon $f_{a/\gamma}(x_\gamma,M_{\gamma})$, can
be found in \cite{Klasen:2005dq}. The main term is proportional to
$\ln(M_{\gamma}^2/Q^2)$ times the splitting function
\beq
 P_{q_i \leftarrow \gamma}(z) = 2 N_c Q_i^2 \frac{z^2+(1-z)^2}{2},
 \label{eq:2}
\eeq
where $z=p_1p_2/p_0q \in [x;1]$ and $Q_i$ is the fractional charge of the
quark $q_i$. $p_1$ and $p_2$ are the momenta of the two outgoing jets, and
$p_0$ and $q$ are the momenta of the ingoing parton and virtual photon,
respectively. Since $Q^2=-q^2 \ll M_{\gamma}^2$, the subtraction term is
large and is therefore resummed by the DGLAP evolution equations for the
virtual photon PDFs. After this subtraction, the finite term
$M(Q^2)_{\overline{\rm MS}}$, which remains in the matrix element for the
NLO correction to the direct process \cite{Klasen:1997jm}, has the same 
$M_{\gamma}$-dependence as the subtraction term, {\it i.e.} $\ln M_{\gamma}$
is multiplied with the same factor. As already mentioned, this yields the
$M_{\gamma}$-dependence before the evolution is turned on. In the usual
non-diffractive dijet photoproduction these two $M_{\gamma}$-dependences
cancel, when the NLO correction to the direct part is added to the LO
resolved  cross section \cite{BKS}. Then it is obvious that the approximate
$M_{\gamma}$-independence is destroyed, if the resolved cross section is
multiplied by a suppression factor $R$ to account for the factorization
breaking in the experimental data. To remedy this deficiency, we propose to
multiply the $\ln M_{\gamma}$-dependent term in $M(Q^2)_{\overline{\rm MS}}$
with the same suppression factor as the resolved cross section. This is done
in the following way: we split $M(Q^2)_{\overline{\rm MS}}$ into two terms
using the scale $p_T^{*}$ in such a way that the term containing the slicing
parameter $y_s$, which was used to separate the initial-state singular
contribution, remains unsuppressed. In particular, we replace the finite
term after the subtraction by
\bea
 M(Q^2,R)_{\overline{\rm MS}} &=& \le-\frac{1}{2N_c} P_{q_i\leftarrow
 \gamma}(z)\ln\lr\frac{M_{\gamma}^2 z}{p_T^{*2}(1-z)}\rr+{Q_i^2\over2} \re R
 \nonumber \\
 && \ -\frac{1}{2N_c} P_{q_i\leftarrow\gamma}(z)
 \ln\lr\frac{p_T^{*2}}{zQ^2+y_s s}\rr,\label{eq:4}
\eea
where $R$ is the suppression factor. This expression coincides with the
finite term after subtraction (see Ref.\ \cite{Klasen:2005dq}) for $R=1$, as
it should, and leaves the second term in Eq.\ (\ref{eq:4}) unsuppressed. In
Eq.\ (\ref{eq:4}) we have suppressed in addition to $\ln(M_{\gamma}^2/
p_T^{*2})$ also the $z$-dependent term $\ln (z/(1-z))$, which is specific to
the $\overline{\rm MS}$ subtraction scheme as defined in
\cite{Klasen:1997jm}. The second term in Eq.\ (\ref{eq:4}) must be left in
its original form, {\it i.e.} being unsuppressed, in order to achieve the
cancellation of the slicing parameter ($y_s$) dependence of the complete NLO
correction in the limit of very small $Q^2$ or equivalently very large $s$.
It is clear that the suppression of this part of the NLO correction
to the direct cross section will change the full cross section only very
little as long as we choose $M_{\gamma} \simeq p_T^{*}$. The first term in
Eq.\ (\ref{eq:4}), which has the suppression factor $R$, will be denoted by
${\rm DIR}_{\rm IS}$ in the following.

To study the left-over $M_{\gamma}$-dependence of the physical cross
section, we have calculated the diffractive dijet cross section with the
same kinematic constraints as in the H1 experiment \cite{Schatzel:2004be}. 
Jets are defined by the CDF cone algorithm with jet radius equal to one and
asymmetric cuts for the transverse momenta of the two jets required for
infrared stable comparisons with the NLO calculations \cite{Klasen:1995xe}.
The original H1 analysis actually used a symmetric cut of 4 GeV on the
transverse momenta of both jets \cite{Adloff:2000qi}. The data have,
however, been reanalyzed for asymmetric cuts \cite{Schatzel:2004be}. 

For the NLO resolved virtual photon predictions, we have used the PDFs SaS1D
\cite{Schuler:1996fc} and transformed them from the DIS$_{\gamma}$ to the 
$\overline{\rm MS}$ scheme as in Ref.\ \cite{Klasen:1997jm}. If not stated
otherwise, the renormalization and factorization scales at the pomeron and
the photon vertex are equal and fixed to $p_T^{*} = p_{T,jet1}^{*}$. We
include four flavors, {\it i.e.} $n_f=4$ in the formula for $\alpha_s$ and
in the PDFs of the pomeron and the photon. With these assumptions we have
calculated the same cross section as in our previous work
\cite{Klasen:2004ct}. First we investigated how the cross section
$\d\sigma/\d Q^2$ depends on the factorization scheme of the PDFs for the
virtual photon, {\it i.e.} $\d\sigma/\d Q^2$ is calculated for the choice
SaS1D and SaS1M. Here $\d\sigma/\d Q^2$ is the full cross section (sum of
direct and resolved) integrated over the momentum and rapidity ranges as in
the H1 analysis. The results, shown in Fig.\ 2 of Ref.\ \cite{Klasen:2005dq},
demonstrate that the choice of the factorization scheme
of the virtual photon PDFs has negligible influence on $\d\sigma/\d Q^2$
for all considered $Q^2$. The predictions agree reasonably well with the
preliminary H1 data \cite{Schatzel:2004be}. 

\begin{figure}
 \centering
 \includegraphics[width=.4\textwidth]{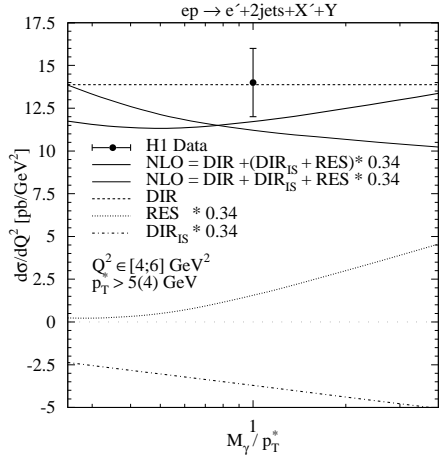}
 \caption{\label{fig:4}Photon factorization scale dependence of resolved
 and direct contributions to $\d\sigma/\d Q^2$ together with their
 weighted sums for (i) suppression of the resolved cross section and for
 (ii) additional suppression of DIR$_{\rm IS}$, using SaS1D virtual
 photon PDFs \cite{Schuler:1996fc}.}
\end{figure}

We now turn to the $M_{\gamma}$-dependence of the cross section with a
suppression factor for DIR$_{\rm IS}$. To show this dependence for the two
suppression mechanisms, (i) suppression of the resolved cross section only
and (ii) additional suppression of the DIR$_{\rm IS}$ term as defined in
Eq.\ (\ref{eq:4}) in the NLO correction of the direct cross section, we
consider $\d\sigma/\d Q^2$ for the lowest $Q^2$-bin, $Q^2\in [4,6]$ GeV$^2$.
In Fig.\ \ref{fig:4}, this cross section
%
%
is plotted as a function of $\xi=M_{\gamma}/p_T^{*}$ in the range $\xi\in
[0.25;4]$ for the cases (i) (light full curve) and (ii) (full curve). We see
that the cross section for case (i) has an appreciable $\xi$-dependence in
the considered $\xi$ range of the order of $40\%$, which is caused by the
suppression of the resolved contribution only. With the additional
suppression of the DIR$_{\rm IS}$ term in the direct NLO correction, the
$\xi$-dependence of $\d\sigma/\d Q^2$ is reduced to approximately less
than $20\%$, if we compare the maximal and the minimal value of $\d\sigma/
\d Q^2$ in the considered $\xi $ range. The remaining $\xi $-dependence is
caused by the NLO corrections to the suppressed resolved cross section and
the evolution of the virtual photon PDFs. How the compensation of the
$M_{\gamma}$-dependence between the suppressed resolved contribution and the
suppressed direct NLO term works in detail is exhibited by the dotted and
dashed-dotted curves in Fig.\ \ref{fig:4}. The suppressed resolved
term increases and the suppressed direct NLO term decreases by approximately
the same amount with increasing $\xi$. In addition we show also $\d\sigma/
\d Q^2$ in the DIS theory, {\it i.e.} without subtraction of any $\ln Q^2$
terms (dashed line). Of course, this cross section must be independent of
$\xi$. This prediction agrees very well with the experimental point, whereas
the result for the subtracted and suppressed theory (full curve) lies
slightly below. We notice, that for $M_{\gamma}=p^{*}_T$ the additional
suppression of DIR$_{\rm IS}$ has only a small effect. It increases
$\d\sigma/\d Q^2$ by $5\%$ only.

\section{Summary}
\label{sec:4}

Experimental data from the H1 and ZEUS Collaborations at HERA for diffractive
dijet production in DIS and photoproduction have been compared with NLO QCD
predictions using diffractive parton densities from H1 and ZEUS. While good
agreement was found for DIS assuming the H1 diffractive PDFs, 
the dijet photoproduction data are overestimated
by the NLO theory, showing that factorization breaking occurs at this order. 
While this is expected theoretically for resolved photoproduction, the fact
that the data are better described by a global suppression of direct {\em and}
resolved contribution by about a factor of two has come as a surprise. We
have therefore discussed in some detail the factorization scheme and scale
dependence between direct and resolved contributions and proposed a new
factorization scheme for diffractive dijet photoproduction.

\section*{Acknowledgments}
M.K.\ thanks the II.\ Institute for Theoretical Physics at the University
of Hamburg for hospitality while this work was being finalized.



\begin{thebibliography}{99}

\bibitem{h1ichep02}
H1 Collaboration,
Abstract 980, contributed to the 31$^{\rm st}$ International Conference
on High Energy Physics (ICHEP 2002), Amsterdam, July 2002.


\bibitem{Chekanov:2004hy}
S. Chekanov et al. [ZEUS Collaboration], \Journal{\EJC}{38}{2004}{43}, 
and A.~Proskuryakov, private communication.

\bibitem{zeus_mx}
M.~Groys, A.~Levy and A.~Proskuryakov, these proceedings.

\bibitem{Schilling}
F.-P.~Schilling and P.~Newman, these proceedings.

\bibitem{Martin:2004xw}
A.~D.~Martin, M.~G.~Ryskin and G.~Watt, Eur.\ Phys.\ J.\ C {\bf37}, 285 (2004)
and DESY 05-055, IPPP/05/07, DCPT/05/14, hep-ph/0504132.

\bibitem{Collins:1997sr}
J.~C.~Collins,
Phys.\ Rev.\ D {\bf 57}, 3051 (1998)
[Erratum-ibid.\ D {\bf 61}, 019902 (2000)].

\bibitem{Affolder:2000vb}
T.~Affolder {\it et al.}  [CDF Collaboration],
Phys.\ Rev.\ Lett.\  {\bf 84}, 5043 (2000).

\bibitem{Kaidalov:2001iz}
A.~B.~Kaidalov, V.~A.~Khoze, A.~D.~Martin and M.~G.~Ryskin,
Eur.\ Phys.\ J.\ C {\bf 21}, 521 (2001).

\bibitem{Klasen:2002xb}
M.~Klasen,
Rev.\ Mod.\ Phys.\  {\bf 74}, 1221 (2002).

\bibitem{Kaidalov:2003xf}
A.~B.~Kaidalov, V.~A.~Khoze, A.~D.~Martin and M.~G.~Ryskin,
Phys.\ Lett.\ B {\bf 567}, 61 (2003).

\bibitem{Klasen:2004tz}
M.~Klasen and G.~Kramer,
hep-ph/0401202,
Proceedings of the 12$^{\rm th}$ International Workshop on Deep Inelastic
Scattering (DIS 2004), eds.\ D.\ Bruncko, J.\ Ferencei and P.\ Strizenec,
Kosice, Inst.\ Exp.\ Phys.\ SAS, 2004, p.\ 492.

\bibitem{Klasen:2004qr}
M.~Klasen and G.~Kramer,
Eur.\ Phys. J.\ C {\bf38}, 93 (2004).

\bibitem{h1ichep04}
H1 Collaboration,
Abstract 6-0177, contributed to the 32$^{\rm nd}$ International Conference
on High Energy Physics (ICHEP 2004), Beijing, August 2004.

\bibitem{zeusichep04}
ZEUS Collaboration,
Abstract 6-0249, contributed to the 32$^{\rm nd}$ International Conference
on High Energy Physics (ICHEP 2004), Beijing, August 2004.

\bibitem{Klasen:2004ct}
M.~Klasen and G.~Kramer,
Phys.\ Rev.\ Lett.\  {\bf 93}, 232002 (2004).

\bibitem{BKS} D.~B\"odeker, G.~Kramer and S.~G.~ Salesch, Z.\ Phys.\ C
{\bf63}, 471 (1994).
 
\bibitem{Klasen:1997jm}
M.~Klasen, G.~Kramer and B.~P\"otter,
Eur.\ Phys.\ J.\ C {\bf 1}, 261 (1998).


\bibitem{disent}
S.~Catani, M.H.~Seymour, \Journal{\NPB}{485}{1997}{291}; erratum ibid. {\bf B510} (1997) 503.

\bibitem{rapgap}
H.~Jung, \Journal{\CPC}{86}{1995}{147}.

\bibitem{zeus_disjets}
ZEUS Collaboration, Abstract 295 and addendum, contributed
to the 22$^{\rm nd}$ 
International Symposium on Lepton-Photon Interactions oh High Energy,
Uppsala, Sweden, June 2005. 

\bibitem{dstar}
H1 Collaboration, Abstract 6-0178, 
contributed to the 32$^{\rm nd}$ International Conference
on High Energy Physics (ICHEP 2004), Beijing, August 2004.


\bibitem{zeus_dstar}
S. Chekanov et al. [ZEUS Collaboration], \Journal{\NPB}{672}{2003}{3}, 

\bibitem{frixione}  S.~Frixione, Z.~Kunszt and A.~Signer,
  \Journal{\NPB}{467}{1996}{399};\\
 S.~Frixione, \Journal{\NPB}{507}{1997}{295}.

\bibitem{zeus_gp}
ZEUS Collaboration, Abstract 293, contributed
to the 22$^{\rm nd}$ 
International Symposium on Lepton-Photon Interactions oh High Energy,
Uppsala, Sweden, June 2005. 

\bibitem{Potter:1999gg}
B.~P\"otter,
Comput.\ Phys.\ Commun.\  {\bf 133}, 105 (2000).

\bibitem{Klasen:2005dq}
M.~Klasen and G.~Kramer,
hep-ph/0506121, accepted for publication in J.\ Phys.\ G.

\bibitem{Schatzel:2004be}
S.~Sch\"atzel,
hep-ex/0408049,
Proceedings of the 12$^{\rm th}$ International Workshop on Deep Inelastic
Scattering (DIS 2004), eds.\ D.\ Bruncko, J.\ Ferencei and P.\ Strizenec,
Kosice, Inst.\ Exp.\ Phys.\ SAS, 2004, p.\ 529;
H1 Collaboration,
Abstract 6-0176, contributed to the 32$^{\rm nd}$ International Conference
on High Energy Physics (ICHEP 2004), Beijing, August 2004.

\bibitem{Klasen:1995xe}
M.~Klasen and G.~Kramer,
Phys.\ Lett.\ B {\bf 366}, 385 (1996).

\bibitem{Adloff:2000qi}
C.~Adloff {\it et al.}  [H1 Collaboration],
Eur.\ Phys.\ J.\ C {\bf 20}, 29 (2001).

\bibitem{Schuler:1996fc}
G.~A.~Schuler and T.~Sj\"ostrand,
Phys.\ Lett.\ B {\bf 376}, 193 (1996).


\end{thebibliography}
\end{document}